\documentclass[iop]{emulateapj}
\usepackage{epsfig}

\def\boldtheta{\bf{\theta}}
\begin{document}
\title{Observations of the WASP--2 System by the APOSTLE Program}
\author{
  Andrew C. ~Becker \altaffilmark{1},
  Praveen Kundurthy \altaffilmark{1},
  Eric Agol \altaffilmark{1},
  Rory Barnes \altaffilmark{1,2},
  Benjamin F. Williams \altaffilmark{1},
  Amy E. Rose \altaffilmark{1}
}
\altaffiltext{1}{Astronomy Department, University of Washington, Seattle, WA 98195}
\altaffiltext{2}{Virtual Planetary Laboratory, USA}

\begin{abstract} 
We present transit observations of the WASP--2 exoplanet system by the Apache Point Survey of Transit Lightcurves of Exoplanets (APOSTLE) program.  Model fitting to these data allows us to improve measurements of the hot--Jupiter exoplanet WASP--2b and its orbital parameters by a factor of $\sim 2$ over prior studies; we do not find evidence for transit depth variations.  We do find reduced $\chi^2$ values greater than 1.0 in the observed minus computed transit times. A sinusoidal fit to the residuals yields a timing semi--amplitude of 32 seconds and a period of 389 days.  However, random rearrangements of the data provide similar quality fits, and we cannot with certainty ascribe the timing variations to mutual exoplanet interactions.  This inconclusive result is consistent with the lack of incontrovertible transit timing variations (TTVs) observed in other hot--Jupiter systems.  This outcome emphasizes that unique recognition of TTVs requires dense sampling of the libration cycle (e.g. continuous observations from space-based platforms).  However, even in systems observed with the Kepler spacecraft, there is a noted lack of transiting companions and TTVs in hot--Jupiter systems.  This result is more meaningful, and indicates that hot--Jupiter systems, while they are easily observable from the ground, do not appear to be currently configured in a manner favorable to the detection of TTVs. The future of ground--based TTV studies may reside in resolving secular trends, and/or implementation at extreme quality observing sites to minimize atmospheric red noise.

\end{abstract}

\section{Introduction}

The transit technique is a highly efficient means of searching for
exoplanetary systems.  The detection of transiting systems requires a
fortuitous observing geometry during the experiment -- the exoplanet
must be observed to traverse its host's stellar disk, occurring for
$~\sim 10\%$ of viewing angles for hot Jupiters, but only $0.4\%$ for
Earth analogues -- and a control of experimental systematics at or
below the level of the transit depth.  For hot--Jupiter systems around
Solar--type stars, this may be as high as $1\%$ ($10^{4}$ parts per
million, ppm) of the out--of--transit depth; an Earth analogue in the
same system would cause a transit depth of only 85 ppm.  Ground--based
observations have been able to achieve per--exposure precisions of
down to 211 ppm \citep{2012arXiv1212.0686T} to 250 ppm
\citep{1993AJ....106.2441G} for the brightest objects.  More commonly,
in transit follow--up efforts where measurements of faint stars in the
field is not a priority, relative photometry at the 300--500 ppm level
is achieved through defocusing and precision tracking, which minimize
the sampling of the flat--field function, and allow the observer to
accumulate more photons before saturation
\citep[e.g.][]{2009MNRAS.396.1023S,2009AJ....137.3826W,2012A&A...542A...4G,2012arXiv1212.3553L}.
Methods to model the non--random (``red'') noise in the data are also
shown to improve the accuracy of photometric transit measurements
\citep{2009ApJ...704...51C}.  However, relative precision in
ground--based data is ultimately limited by atmospheric decoherence
between the target and comparison stars and across observing epochs,
with leading terms including the structure function of clouds
\citep{2007AJ....134..973I}, the time--rate of change of aerosols and
water vapor in the atmosphere \citep{2007PASP..119.1163S}, and
atmospheric scintillation \citep{1991PASP..103..221Y}.  Space--based
observations by the Kepler spacecraft \citep{2010Sci...327..977B} set
the gold--standard for relative photometry, reaching
root--mean--square systematic variations of 20 ppm on timescales of
several hours \citep{2011ApJS..197....6G}.

As recognized by \cite{2005MNRAS.359..567A} and
\cite{2005Sci...307.1288H}, the times of transits in multi--planet
systems may not be exactly periodic due to mutual gravitational
interactions of the planets.  Multiple studies have been undertaken to
search for these transit timing variations (TTVs) in known transiting
exoplanet systems, using high--precision follow--up observations.  The
vast majority of follow--up has been undertaken on systems originally
discovered from the ground, which have been heavily biased towards
hot--Jupiter systems.  The follow--up sampling of these transits is
often irregular, due to weather, daytime, and seasonal effects.  This
makes detection of the libration of transit times, expected to occur
over the timescales of months to years, difficult to recognize because
the signal is undersampled.
\cite{2012MNRAS.422L..57B} outline the difficulties in resolving
TTVs in Jovian systems: they are mostly relevant for systems near (but
not exactly at) mean--motion resonance, and even detected signals may
yield degenerate solutions for the mass of the perturber.
No unambiguous TTVs have yet been discovered in ground--based
follow--up \citep[see however][]{2011A&A...536L...9T}.  In contrast,
the Kepler spacecraft follows--up its own discoveries through
continuous lightcurve coverage.  Several multi--planet systems have
been discovered through Kepler TTVs
\citep[e.g.][]{2012ApJ...750..113F,2012MNRAS.421.2342S,2012ApJ...750..114F},
which have proven to be a powerful verification and mass measurement
technique for planet candidates \citep{2011ApJS..197....7C}.

\section{APOSTLE Program}

The Apache Point Survey of Transit Lightcurves of Exoplanets
\citep[APOSTLE;][]{praveen-tres3} program was initiated as a
systematic study of known transiting exoplanet systems on the ARC 3.5m
telescope + {\tt Agile} imager \citep{2011PASP..123.1423M}.  The large
aperture of the system and frame--transfer capabilities of {\tt Agile}
allow us to obtain high--precision (500 ppm RMS) relative photometry
between $R = 10.8^{th}$ and $R = 10.8^{th}$ magnitude stars
\citep[XO--2;][]{praveen-xo2}, decreasing to 700 ppm at $R = 12.2$
vs. $R = 12.9$ \citep[TrES--3;][]{praveen-tres3}, and 1000 ppm at $R =
13.8$ vs. $R = 13.6$ \citep[GJ 1214;][]{2011ApJ...731..123K}.
Importantly, these observations happen at 100\% duty cycle due to {\tt
  Agile}'s frame--transfer capabilities.  We make use of the {\tt
  MultiTransitQuick} modeling program described in
\cite{praveen-tres3}, which uses a Markov Chain Monte Carlo (MCMC)
analyzer alongside a lightcurve parameterization that minimizes
degeneracies between fitted parameters, ensuring that the MCMC
proceeds efficiently and faithfully samples parameter space.

This paper describes APOSTLE observations of the WASP--2
\citep{2007MNRAS.375..951C} system.  The host star WASP--2A is a $R =
11.3$, spectral type K1 dwarf, with an effective temperature of
T$_{\text{eff}} = 5110 \pm 60$ inferred from optical and infrared
colors \citep{2011MNRAS.418.1039M}, T$_{\text{eff}} = 5150 \pm 80 K$
using photospheric fitting of spectroscopic data
\citep{2010A&A...524A..25T}, and metallicity of $[Fe/H] = −0.08 \pm
0.08$ \citep{2010A&A...524A..25T}.  In our modeling we included
photometric data from previous publications including
\cite{2010MNRAS.408.1680S} -- who converted the timings from
\cite{2007MNRAS.375..951C}, \cite{2007ApJ...658.1322C},
\cite{2009IAUS..253..446H} into the common time standard BJD(TDB) as
outlined by \cite{2010PASP..122..935E} -- as well as one transit epoch
from \cite{2012PASP..124..212S}.

\section{APOSTLE Observations of WASP--2}

\begin{figure}[t]
\centering
\includegraphics[width=0.49\textwidth]{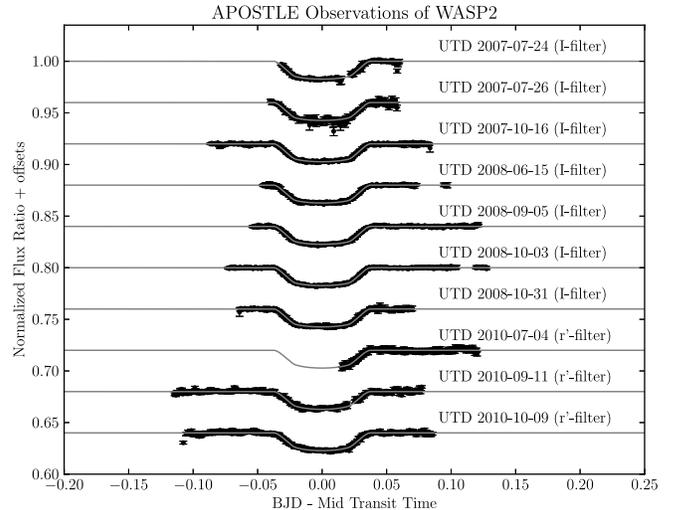}
\caption{
Seven I-–band and three r-–band detrended lightcurves of the WASP-2 system.  The
vertical axis is in normalized flux ratio units. The horizontal axis
shows time from the mid--transit time in days, computed by subtracting
the appropriate mid--transit time for each transit from the best--fit
values in the $\boldtheta_{\rm Multi-Depth}$ chain.
}
\label{fig:lcs}
\end{figure}

The APOSTLE data reduction pipeline is described in prior
publications, including our observational techniques, details of
photometric extraction, detrending of the lightcurves, and
parameterization of the transit model \citep{praveen-tres3}.
Specifically to the WASP--2 system, we acquired 7 Cousins $I$--band
and 3 $r$--band transit sequences between July 2007 and October 2010.
The $I$--band images were fringe--corrected using the techniques
described in \cite{praveen-xo2}.  Observations were taken with a
variety of instrumental settings, staring with 0.5s observations on
the nights of 2007-07-24 and 2007-07-26, and moving to longer
defocussed exposures starting in 2010.  For the analysis here, all
data were binned to equivalent--45s exposures.  For all observations
we used the $R = 11.4$ comparison star TYC 522-780-1.

\section{Model Fitting}

Reduced lightcurves were detrended and modeled using the {\tt
  MultiTransitQuick (MTQ)} package \citep{praveen-tres3}.  We used
{\tt MTQ} in two modes: to fit for similar transit depths for data
taken in a given filter (Multi-Filter model); and to fit each transit
depth individually (Multi-Depth).  The set of parameters used for
Multi--Filter version of {\tt MTQ} is $\boldtheta_{\rm Multi-Filter} =
\{t_T, t_G, D_{j...N_F}, v_{1,j...N_F}, v_{2,j...N_F}, T_{i...N_T}\}$,
where $t_T$ is transit duration and $t_G$ the limb--crossing duration.
The per--filter fit parameters, up to the number of filters $N_F$,
include the transit depth $D$, and limb darkening parameters $v_1$ and
$v_2$.  Finally, the parameter set includes mid--transit times $T_i$
up to the number of transits $N_T$.  For the Multi--Depth version we
used $\boldtheta_{\rm Multi-Depth} = \{t_T, t_G, D_{i...N_T},
v_{1,j...N_F}, v_{2,j...N_F}, T_{i...N_T}\}$, the main difference
being we fit for each transit depth separately ($N_T$) instead of
per--filter ($N_F$).  

As outlined in \cite{praveen-tres3} and \cite{praveen-xo2}, we did not
fit for the limb darkening coefficients as they result in poorly
converged Markov chains.  Instead we kept them fixed at values
determined using the \cite{2011A&A...529A..75C} quadratic limb
darkening models: $u_{1,I} = 0.3926$, $u_{2,I} = 0.2166$, $u_{1,r} =
0.5541$, $u_{2,r} = 0.1594$ \citep[see][for cautions regarding this
  procedure]{2013A&A...549A...9C}.  Our limb darkening terms $v_1,v_2$
are linear combinations of the \cite{2011A&A...529A..75C} quadratic
terms, $v_1 = u_1 + u_2$ and $v_2 = u_1 - u_2$.

Figure~\ref{fig:lcs} presents our detrended data, offset for clarity,
with the best--fit model lightcurves overplotted.  The
root--mean--square scatter about the model fits ranged from 472 ppm
(2008-10-03) to 2252 ppm (2007-07-26), with a median of 626 ppm for
the $I$--band data, and 1146 for the $r$--band data.

High resolution imaging of the system by \cite{2009A&A...498..567D}
reveals a faint companion within the wings of the WASP--2 host star,
which will affect our conversion from $D$ to $R_p/R_\star$.  This
required that we estimate the apparent $I$--band and $r$--band
magnitudes of WASP--2A and the contaminating star (named here
WASP--2/C).

We converted the reported $i'$ and $z'$ magnitudes from
\cite{2009A&A...498..567D} using the transformations presented in
\cite{2006AJ....132..989R} and \cite{2006A&A...460..339J}, as well as
on the SDSS DR7 webpage
\footnote{http://www.sdss.org/dr7/algorithms/jeg\_photometric\_eq\_dr1.html}.
The final estimates yielded $r_{2A} = 11.68 \pm 0.11$, $I_{2A} = 11.00
\pm 0.11$, $r_{2/C} = 17.05 \pm 0.20$, $I_{2/C} = 15.39 \pm 0.20$.
The two dominant uncertainties in these conversions are: the apparent
brightnesses reported by \cite{2009A&A...498..567D}, which are
uncertain to 0.1 magnitudes; and for WASP--2/C its redder color, which
leads to a larger uncertainty in each color--term.
This result indicated that $0.7\%$ of the stellar flux in the
$r$--band comes from WASP--2/C, and $1.7\%$ in the $I$--band; the
WASP--2b transit depth $D$ increased proportionally.  The change in
the $I$--band depth was approximately four times the parameter
uncertainty determined below, making this a necessary correction.  The
derived values of $R_p/R_\star$ increased by $0.3\%$ and $0.8\%$ in
the $r$--band and $I$--band, respectively.

For each parameter set ($\boldtheta_{\rm Multi-Filter}$ and
$\boldtheta_{\rm Multi-Depth}$), we ran two MCMC chains, each having
$2 \times 10^6$ steps.  These were cropped at the beginning of the
chains, where the step acceptance rate is lower than optimal
\citep{gelman2003}, yielding approximately $1.8 \times 10^6$ steps per
chain used in the subsequent analysis.  These chains were compared
against each other to evaluate the Gelman--Rubin $\hat{\rm R}$--static
\citep{Gelman92} and assure that the chains sufficiently sampled model
space.

Evaluation of our Multi--Depth fit indicated that the depths in the
$I$--band were consistent at $D_I = 0.0174 \pm 0.0003$ magnitudes,
while in the $r$--band the depth were also consistent ($D_r = 0.017
\pm 0.001$) but with a larger RMS due to having only 2 completely
sampled transits\footnote{Note that these depths were determined
  before the corrections for WASP--2/C were applied, and thus differ
  from the final Multi--Filter results presented in
  Table~\ref{tab-pars}}.  We found no evidence for transit depth
variations within these data, and present the Multi--Filter fits as
our final results.

\begin{figure*}[t] 
\begin{center} 
\epsfig{file=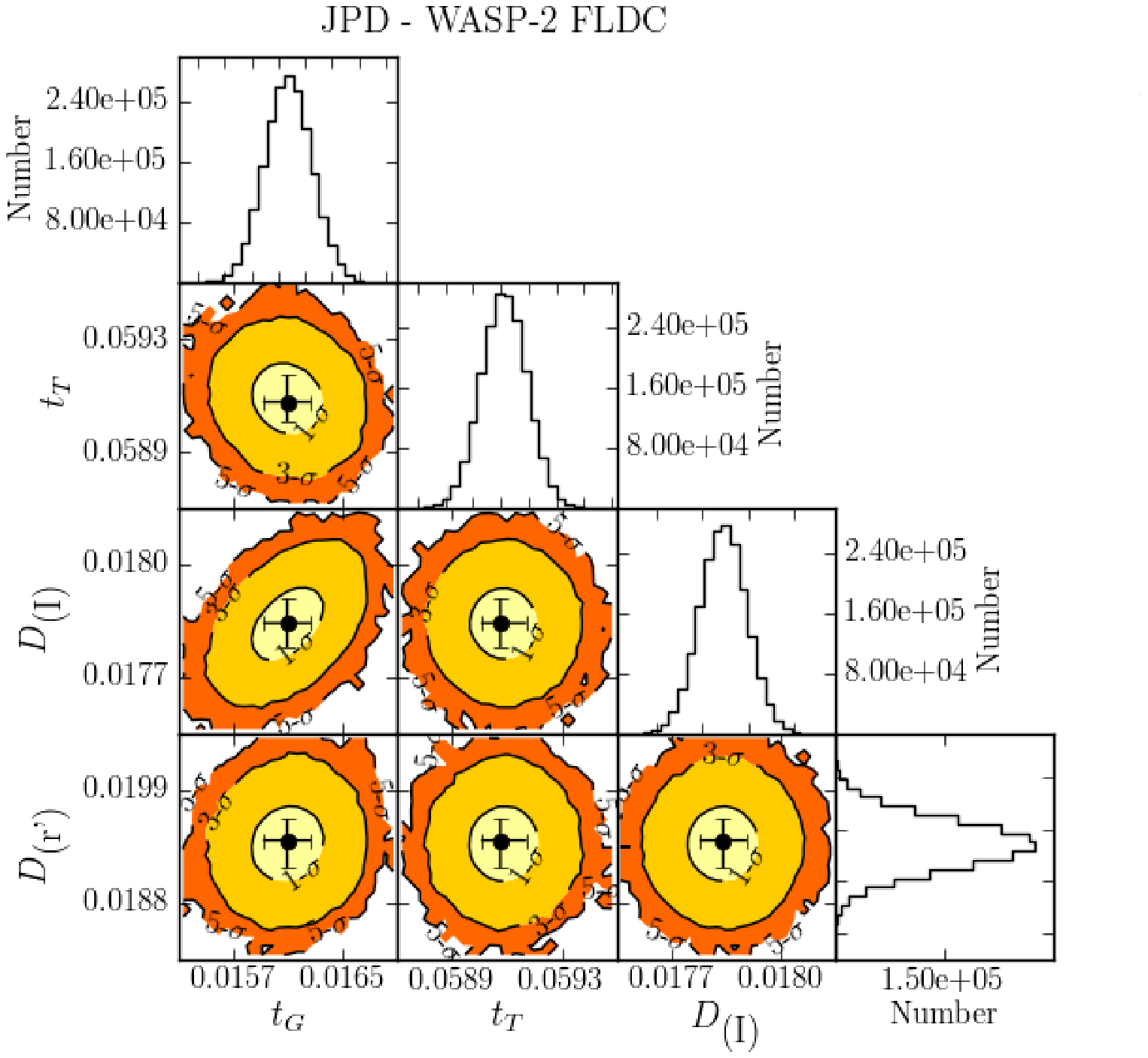, width=0.49\textwidth} 
\epsfig{file=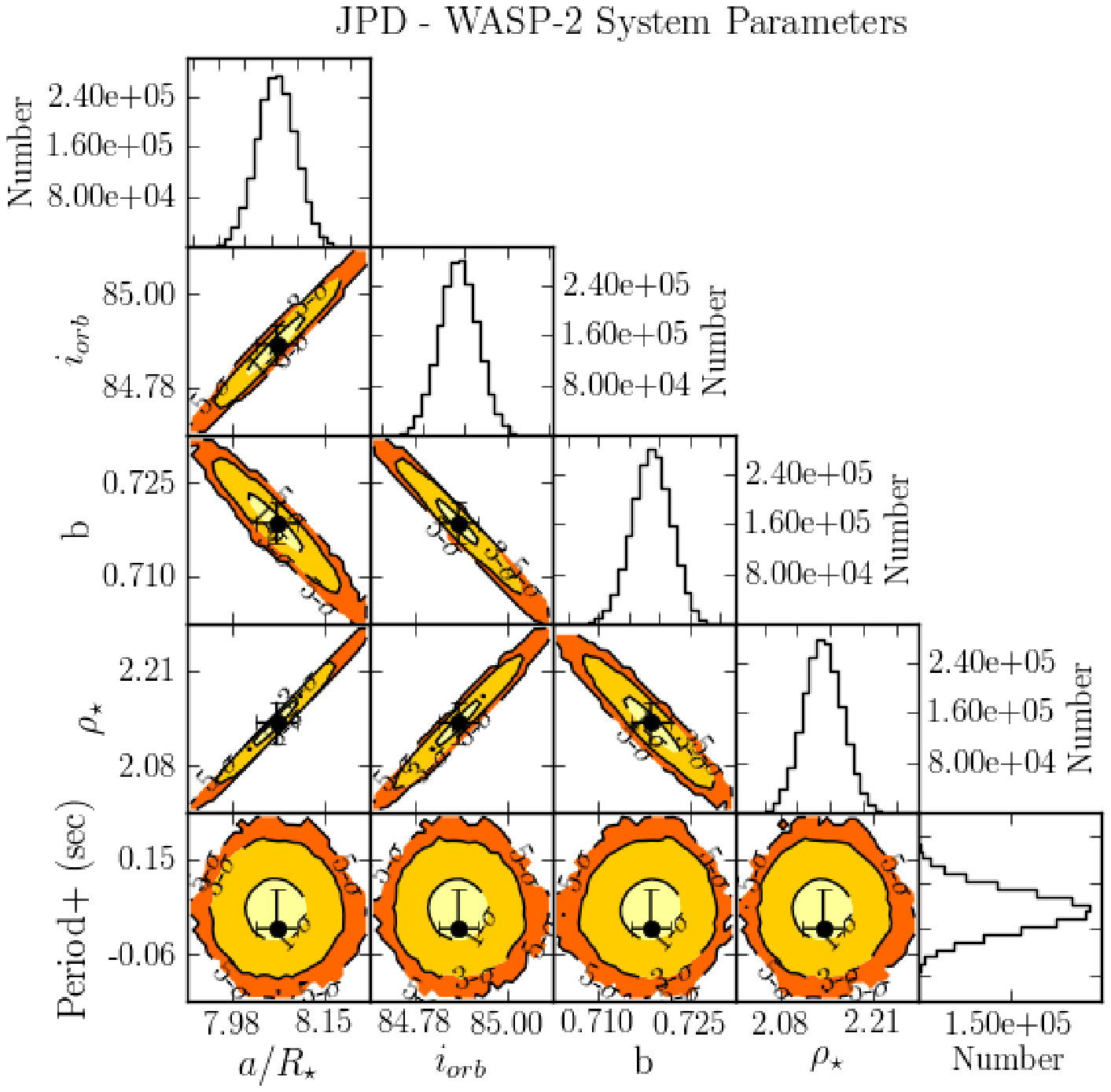, width=0.49\textwidth} 
\caption{Plots of the joint probability distributions (JPD) of
  parameters from the Multi--Filter chains with fixed limb darkening.
  The {\it left} panel shows the fitted--for parameters, which are
  weakly correlated.  The {\it right} panels shows the derived
  parameters, which tend to show larger correlations.
  Table~\ref{tab-pars} provides the relevant units for each parameter
  set.}
\label{fig-jpd} 
\end{center} 
\end{figure*}

Fitted--for and derived Multi--Filter parameters are presented in
Table~\ref{tab-pars}, with the joint--probability distributions
presented in Figure~\ref{fig-jpd} for the fitted and derived
parameters.  Derived parameters include: $R_{p}/R_{\star}$, the radius
of the planet in units of the host stellar radius; $a/R_{\star}$, the
normalized semi--major axis $a$ of the planetary orbit; the stellar
density $\rho_{\star}$; planet impact parameter $b$; orbital
inclination $i$; and orbital period $P$.  We found values of the
Gelman--Rubin $\hat{\rm R}$--static within $10^{-3}$ of 1.0 for all
fitted parameters, indicating sufficient coverage of the chain over
parameter space.  The shortest effective chain length is for the time
of transit on 2007-07-26 with a length of 9419.  This night had some
of the largest photometric uncertainties, and largest overall
uncertainty on the time of transit.  All other parameters have
effective chain lengths larger than $10^4$, indicative of sufficient
mixing in the MCMC sample \cite[e.g.][]{2004PhRvD..69j3501T}.
Finally, the $\chi^2$ of the best model fit to the data is 2822.93 for
2809 degrees of freedom (reduced $\chi^2$ of 1.005), indicating that
our data (and therefore parameter) uncertainties are well understood.

\section{System Parameters}

Fitted--for and derived system parameters are presented in
Table~\ref{tab-pars}, along with their uncertainties.  We note that
the reported transit depths are for the WASP--2 system only, after
correction for WASP--2/C.  For comparison, we present the results from
\citet[][S10]{2010MNRAS.408.1680S}, who observed in Cousins $R$--band.
Figure~\ref{fig-jpd} presents the joint probability distributions of
fitted--for ({\it left}) and derived parameters ({\it right}).  The
left panel shows a much smaller level of correlation between
parameters, making it an effective basis in which to perform MCMC.
Final {\tt MTQ} times of transits for each night of observation are
presented in the left--hand columns of Table~\ref{table-tt}.

To model the effects of correlated lightcurve noise on our transit
times, we used the Transit Analysis Package \citep[{\tt
    TAP};][]{2012AdAst2012E..30G}, which is an implementation of the
red--noise model of \cite{2009ApJ...704...51C}.  Importantly, {\tt
  TAP} models the amplitude of ``white'' (random) and ``red''
(correlated) noise on each night of observations.  We applied {\tt
  TAP} using the period derived from {\tt MTQ}, fixed limb darkening
values, and zero eccentricity and argument of periastron.  For APOSTLE
observations of WASP--2, the median red (white) noise contributions
were 0.002 (0.0006) magnitudes in the $I$--band, and 0.006 (0.0006)
magnitudes in the $r$--band.  This indicated that while the
statistical noise in the two datasets is comparable, the correlated
noise dominates, and is larger in the $r$--band data.  This suggests
that the correlated noise comes from intrinsic stellar variability
(either of the target or comparison star), or perhaps that the nights
of $r$--band observation happened to have larger time--variation in
atmospheric molecular water absorption \citep{2007PASP..119.1163S}.
The {\tt TAP} system parameters are included in
Table~\ref{tab-pars}, and times of transit in the right--hand
columns of Table~\ref{table-tt}.  While the {\tt TAP} uncertainties
are typically larger than those from {\tt MTQ}, in all cases the
APOSTLE analysis yields an improvement in precision over previous
measurements.

\begin{table*}[t]
\small
\begin{center}
\caption{\label{tab-pars} WASP-2 System Parameters}
\begin{tabular}{ccccc}
\hline \hline
Parameter & Value & TAP & S10 & Unit\\
\hline
\multicolumn{5}{c}{MTQ $\boldtheta_{\text{Multi-Filter}}$ Parameters} \\
\hline
$t_{G}$             &0.0161$\pm$0.0002&\nodata&\nodata&days\\
$t_{T}$             &0.0591$\pm$0.0001&\nodata&\nodata&days\\
$D_{\textrm{(I)}}$  &0.0178$\pm$0.0001&\nodata&\nodata&-\\
$D_{\textrm{(r')}}$ &0.0194$^{+0.0002}_{-0.0003}$&\nodata&\nodata&-\\
$v_1{\textrm{(I)}}$ &(0.6092)         &\nodata&\nodata&-\\
$v_2{\textrm{(I)}}$ &(0.1760)         &\nodata&\nodata&-\\
$v_1{\textrm{(r')}}$&(0.7135)         &\nodata&\nodata&-\\
$v_2{\textrm{(r')}}$&(0.3947)         &\nodata&\nodata&-\\
\hline
\multicolumn{5}{c}{Derived Parameters} \\
\hline
$(R_{p}/R_{\star})_{\textrm{(I)}}$&0.1315$\pm$0.0003            &0.1317$\pm$0.0004&\nodata&-\\
$(R_{p}/R_{\star})_{\textrm{(r')}}$&0.1362$^{+0.0007}_{-0.0009}$&0.1359$\pm$0.001&\nodata&-\\
$(R_{p}/R_{\star})_{\textrm{(R)}}$&\nodata&\nodata&$0.1326 \pm 0.0007$&-\\
$a/R_{\star}$&8.06$\pm$0.04                                     &7.99$\pm$0.06&$8.08 \pm 0.12$&-\\
%$\nu/R_{\star}$&23.53$\pm$0.11&&days$^{-1}$\\
$\rho_{\star}$&2.14$\pm$0.03                                    &2.08$\pm$0.05&2.15$\pm$0.09&g/cc\\
b&0.719$\pm$0.003                                               &0.723$\pm$0.005&\nodata&-\\
i&84.89$\pm$0.05                                                &84.81$\pm$0.08& $84.81 \pm 0.17$ &$^{o} (deg)$\\
P (2.1522 days +)&1812$\pm$26                                   &\nodata&1852$\pm$34&milli-sec\\
\hline
\end{tabular}
\end{center}
\end{table*}

\begin{table}[t]
\begin{center}
\caption{\label{table-tt} APOSTLE Transit Times for WASP2}
\begin{tabular}{ccc|cc}
\hline \hline
Epoch & T0 (MTQ) & $\sigma_{T0}$ & T0 (TAP) & $\sigma_{T0}$ \\
 & 2,400,000+ (BJD) & (BJD) & 2,400,000+ (BJD) & (BJD) \\
\hline
146&54305.73863&0.00035&54305.73862&0.00019\\ 
147&54307.89212&0.00063&54307.89200&0.00065\\ 
185&54389.67652&0.00018&54389.67646&0.00024\\ 
298&54632.87724&0.00010&54632.87720&0.00019\\ 
336&54714.66135&0.00014&54714.66158&0.00018\\ 
349&54742.64007&0.00006&54742.64011&0.00009\\ 
362&54770.61909&0.00008&54770.61897&0.00016\\ 
646&55381.84717&0.00092&55381.84757&0.00072\\ 
678&55450.72118&0.00024&55450.72108&0.00029\\ 
691&55478.70099&0.00025&55478.70063&0.00044\\ 
\hline
\end{tabular}
\end{center}
\end{table}

\subsection{Transit Timing Analysis \label{sec-tt}}

\begin{figure}[t]
\centering
\includegraphics[width=0.49\textwidth]{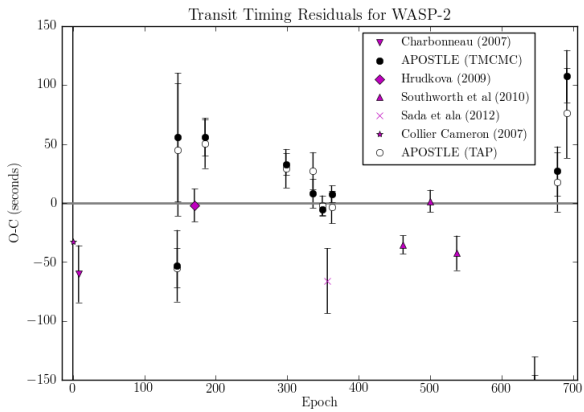}
\caption{
The observed minus computed transit times for WASP--2b. Values from
APOSTLE’s {\tt MTQ} analysis, {\tt TAP}, and previous literature are
plotted.  The horizontal axis represents the transit Epoch. The
zero-line ephemeris is described in Section~\ref{sec-tt}.
}
\label{fig:OC}
\end{figure}

Using the above analysis, we found a revised ephemeris for WASP--2b of
\begin{eqnarray*}
P = 2.152220976 \pm 0.000000305 & \text{days} \\
T0= 2453991.5148944 \pm 0.0001232 & \text{BJD} 
\end{eqnarray*}
using APOSTLE results combined with \cite{2012PASP..124..212S} and the
non--amateur results presented in \cite{2010MNRAS.408.1680S}, for 17
epochs overall.  As Figure~\ref{fig:OC} indicates, there is large
scatter in the observed minus computed transit times (O--C diagram),
with a reduced $\chi^2$ of 4.7 for the {\tt TAP} results (7.0 for {\tt
  MTQ}), but it is difficult to claim a detection of coherent transit
timing variations due to the sparse sampling.

To quantify the significance of this signal, we performed a
3--parameter sinusoidal fit (period in days, amplitude in seconds, and
phase offset as a nuisance parameter) to the O--C data for both the
{\tt TAP} (and {\tt MTQ}) timings.  This yielded a $\Delta \chi^2$
improvement of $28.2$ $(47.7)$, with amplitudes of 32 (34) seconds and
periods of 389 (437) days.  We next performed $10^5$ random
reassignments of the timing data to the epochs of observations
(i.e. we kept the APOSTLE time sampling of any putative TTV signal,
but shuffled the observed amplitudes).  The fitter was initialized to
the period corresponding to the peak of the shuffled--TTV periodogram,
determined using the method of \cite{2009A&A...496..577Z}.  We then
examined what fraction of these random assignments allowed a $\Delta
\chi^2$ equal to or larger than that observed.  This provided an
estimate of the false alarm probability for any potential transit
timing modulation.  We found that $42\%$ ($9.7\%$) of the random
shuffles yielded $\Delta \chi^2$ improvements at an amplitude equal to
or larger than that observed.  While the {\tt MTQ} results were
significant at the $1.7~\sigma$ level, the TAP results are more
realistic given the correlated noise in our data.  Thus, these data
provided only marginal evidence for transit timing variations in the
WASP--2 system.

\section{Results and Conclusions}

We presented observations and analysis of 10 transit timings of the
WASP--2 system by the APOSTLE program.  After an extensive treatment
of the data to address and model the intrinsic systematic errors, we
undertook a MCMC modeling analysis to understand the parameter
uncertainties and correlations.  As in previous publications, a model
analysis incorporating transit duration $t_T$, limb--crossing duration
$t_G$, and transit depth $D$ yielded weakly correlated parameters
(Figure~\ref{fig-jpd}).  We corrected for a previously reported object
within the photometric aperture of the WASP--2b host star
\citep{2009A&A...498..567D} to yield the system parameters reported in
Table~\ref{tab-pars}.  The uncertainties on these parameters are
smaller than those reported in previous studies
\citep{2010MNRAS.408.1680S}.  We disentangled the random and
correlated noise in each lightcurve using the method of
\cite{2012AdAst2012E..30G}, and found that the red noise dominates
scatter in the lightcurves, at a larger amplitude in the $r$--band
than in the $I$--band.

The depths of transit coming from this analysis did not show
significant time dependence.  However, the times of transit shown in
Figure~\ref{fig:OC} show scatter larger than the uncertainties.  While
a basic sinusoidal fit to this signal provided a significant
improvement in $\chi^2$, we achieved a similar goodness of fit in
$18\%$ of random reassignments of the transit timing variations to the
APOSTLE epochs.  We therefore cannot conclusively report evidence of
transit timing variations in the WASP--2 system.

These results mirror many of those reported in the field of
ground--based transit timing measurements, where the O--C diagrams
show reduced $\chi^2$ larger than 1.0, but there is not
incontrovertible evidence of coherent transit timing librations
\citep[e.g.][]{2012MNRAS.423.1381E,2012PASP..124..212S,2012ApJ...748...22H}.
A primary reason for this is sparse sampling of transit epochs, due to
weather, daytime, and seasonal considerations.  In contrast, the
continuous observations afforded by the Kepler spacecraft provide
complete sampling of the TTV signal, with both high signal--to--noise
per transit and a large number of transits observed.  Resolved Kepler
TTVs have sufficient overall signal--to--noise to verify and weigh
planets in the system.  Even with tight control of experimental
systematics, ground--based observations are intrinsically limited by
the spatial and temporal stability of the Earth's atmosphere;
metrology at the level required to achieve calibration at the level of
Kepler is currently not available.

%%%%%%%%%%%%%%

It is likely that the lack of TTVs being detected from the ground is
due to a fundamental property of the systems that are being studied.
The vast majority of ground--based follow--up is focused on
hot--Jupiter systems, since resolving their transit depths is
achievable even with modest aperture telescopes and non--photometric
observing conditions.  However, as outlined by
\cite{2012PNAS..109.7982S}, hot--Jupiter systems observed by Kepler
     {\it also} fail to show detectable TTVs, or any evidence for
     being in a multi--planet system.  Their neighbors in exoplanet
     parameter space, loosely termed warm--Jupiters and hot--Neptunes,
     do show evidence for both TTVs and transiting companions.  What
     this suggests is a unique dynamical pathway for the formation of
     contemporary hot--Jupiters, such as multi--planet scattering
     \cite[e.g.][]{2012ApJ...751..119B} leading to the ejection of
     lesser bodies from the system.  It is thus possible that those
     systems that are most easily observable from the ground are also
     those that will fail to exhibit TTVs.  Given the longevity of
     ground--based observing resources compared to space--based ones,
     it may benefit the field to undertake longer--term observations
     of systems to resolve secular trends such as those arising from
     stellar binary hosts \citep{2000ApJ...535..385F} or tidal orbital
     decay \citep{1981A&A....99..126H}.
     The resolution of TTVs from
     the ground may also require observations from sites designed for the
     minimization of aerosols and water vapor, such as Llano de
     Chajnantor in the Atacama desert of Chile, to reduce atmospheric
     red noise and make resolving transits of other classes of exoplanet 
     systems feasible.

While efforts to resolve TTVs from the ground have been largely
unsuccessful, the drive to follow--up detected exoplanet systems has
enabled significant leaps in the understanding and application of
high--precision relative photometry.  These advances are now extending
to the broad application of time--domain spectroscopy during transits
to map out exoplanet features in absorption
\citep[e.g.][]{2011ApJ...736..132C}, and have overall been a boon to
the field of time--domain astronomy.

\section*{Acknowledgments}
Based on observations obtained with the Apache Point Observatory
3.5-meter telescope, which is owned and operated by the Astrophysical
Research Consortium.  Funding for this work came from NASA Origins
grant NNX09AB32G and NSF Career grant 0645416.  RB acknowledges
funding from the NASA Astrobiology Institute's Virtual Planetary
Laboratory lead team, supported by NASA under cooperative agreement
NNH05ZDA001C.  We would like to thank the APO Staff, APO Engineers,
and R. Owen for helping the APOSTLE program with its observations and
A. Mukadam for instrument characterization.  We would like thank
S. L. Hawley for scheduling our observations on APO.  We thank N.
Thomas for her help in the early analysis of the WASP--2 data.  This
work acknowledges the use of parts of J. Eastman's EXOFAST transit
code as part of APOSTLE's transit model MTQ.

%\bibliographystyle{apj}
%\bibliography{refs,acb_refereed}

\begin{thebibliography}{}

\bibitem[\protect\citeauthoryear{{Agol} et~al.}{{Agol}
  et~al.}{2005}]{2005MNRAS.359..567A}
{Agol}, E., {Steffen}, J., {Sari}, R.,  \& {Clarkson}, W. 2005, \mnras, 359,
  567

\bibitem[\protect\citeauthoryear{{Beaug{\'e}} \& {Nesvorn{\'y}}}{{Beaug{\'e}}
  \& {Nesvorn{\'y}}}{2012}]{2012ApJ...751..119B}
{Beaug{\'e}}, C.,  \& {Nesvorn{\'y}}, D. 2012, \apj, 751, 119

\bibitem[\protect\citeauthoryear{{Borucki} et~al.}{{Borucki}
  et~al.}{2010}]{2010Sci...327..977B}
{Borucki}, W.~J., et~al. 2010, Science, 327, 977

\bibitem[\protect\citeauthoryear{{Bou{\'e}} et~al.}{{Bou{\'e}}
  et~al.}{2012}]{2012MNRAS.422L..57B}
{Bou{\'e}}, G., {Oshagh}, M., {Montalto}, M.,  \& {Santos}, N.~C. 2012, \mnras,
  422, L57

\bibitem[\protect\citeauthoryear{{Cameron} et~al.}{{Cameron}
  et~al.}{2007}]{2007MNRAS.375..951C}
{Cameron}, A.~C., et~al. 2007, \mnras, 375, 951

\bibitem[\protect\citeauthoryear{{Carter} \& {Winn}}{{Carter} \&
  {Winn}}{2009}]{2009ApJ...704...51C}
{Carter}, J.~A.,  \& {Winn}, J.~N. 2009, \apj, 704, 51

\bibitem[\protect\citeauthoryear{{Charbonneau} et~al.}{{Charbonneau}
  et~al.}{2007}]{2007ApJ...658.1322C}
{Charbonneau}, D., {Winn}, J.~N., {Everett}, M.~E., {Latham}, D.~W., {Holman},
  M.~J., {Esquerdo}, G.~A.,  \& {O'Donovan}, F.~T. 2007, \apj, 658, 1322

\bibitem[\protect\citeauthoryear{{Claret} \& {Bloemen}}{{Claret} \&
  {Bloemen}}{2011}]{2011A&A...529A..75C}
{Claret}, A.,  \& {Bloemen}, S. 2011, \aap, 529, A75

\bibitem[\protect\citeauthoryear{{Cochran} et~al.}{{Cochran}
  et~al.}{2011}]{2011ApJS..197....7C}
{Cochran}, W.~D., et~al. 2011, \apjs, 197, 7

\bibitem[\protect\citeauthoryear{{Crossfield}, {Barman}, \&
  {Hansen}}{{Crossfield} et~al.}{2011}]{2011ApJ...736..132C}
{Crossfield}, I.~J.~M., {Barman}, T.,  \& {Hansen}, B.~M.~S. 2011, \apj, 736,
  132

\bibitem[\protect\citeauthoryear{{Csizmadia} et~al.}{{Csizmadia}
  et~al.}{2013}]{2013A&A...549A...9C}
{Csizmadia}, S., {Pasternacki}, T., {Dreyer}, C., {Cabrera}, J., {Erikson}, A.,
   \& {Rauer}, H. 2013, \aap, 549, A9

\bibitem[\protect\citeauthoryear{{Daemgen} et~al.}{{Daemgen}
  et~al.}{2009}]{2009A&A...498..567D}
{Daemgen}, S., {Hormuth}, F., {Brandner}, W., {Bergfors}, C., {Janson}, M.,
  {Hippler}, S.,  \& {Henning}, T. 2009, \aap, 498, 567

\bibitem[\protect\citeauthoryear{{Eastman}, {Siverd}, \& {Gaudi}}{{Eastman}
  et~al.}{2010}]{2010PASP..122..935E}
{Eastman}, J., {Siverd}, R.,  \& {Gaudi}, B.~S. 2010, \pasp, 122, 935

\bibitem[\protect\citeauthoryear{{Eibe} et~al.}{{Eibe}
  et~al.}{2012}]{2012MNRAS.423.1381E}
{Eibe}, M.~T., {Cuesta}, L., {Ull{\'a}n}, A., {P{\'e}rez-Verde}, A.,  \&
  {Navas}, J. 2012, \mnras, 423, 1381

\bibitem[\protect\citeauthoryear{{Fabrycky} et~al.}{{Fabrycky}
  et~al.}{2012}]{2012ApJ...750..114F}
{Fabrycky}, D.~C., et~al. 2012, \apj, 750, 114

\bibitem[\protect\citeauthoryear{{Ford} et~al.}{{Ford}
  et~al.}{2012}]{2012ApJ...750..113F}
{Ford}, E.~B., et~al. 2012, \apj, 750, 113

\bibitem[\protect\citeauthoryear{{Ford}, {Kozinsky}, \& {Rasio}}{{Ford}
  et~al.}{2000}]{2000ApJ...535..385F}
{Ford}, E.~B., {Kozinsky}, B.,  \& {Rasio}, F.~A. 2000, \apj, 535, 385

\bibitem[\protect\citeauthoryear{{Gazak} et~al.}{{Gazak}
  et~al.}{2012}]{2012AdAst2012E..30G}
{Gazak}, J.~Z., {Johnson}, J.~A., {Tonry}, J., {Dragomir}, D., {Eastman}, J.,
  {Mann}, A.~W.,  \& {Agol}, E. 2012, Advances in Astronomy, 2012

\bibitem[\protect\citeauthoryear{Gelman et~al.}{Gelman
  et~al.}{2003}]{gelman2003}
Gelman, A., Carlin, J.~B., Stern, H.~S.,  \& Rubin, D.~B. 2003, Bayesian Data
  Analysis, Second Edition (Chapman \& {Hall/CRC} Texts in Statistical Science)
  (2 ed.) (Chapman and Hall/CRC)

\bibitem[\protect\citeauthoryear{Gelman \& Rubin}{Gelman \&
  Rubin}{1992}]{Gelman92}
Gelman, A.,  \& Rubin, D. 1992, Statistical Science, 7, 457

\bibitem[\protect\citeauthoryear{{Gilliland} et~al.}{{Gilliland}
  et~al.}{1993}]{1993AJ....106.2441G}
{Gilliland}, R.~L., et~al. 1993, \aj, 106, 2441

\bibitem[\protect\citeauthoryear{{Gilliland} et~al.}{{Gilliland}
  et~al.}{2011}]{2011ApJS..197....6G}
{Gilliland}, R.~L., et~al. 2011, \apjs, 197, 6

\bibitem[\protect\citeauthoryear{{Gillon} et~al.}{{Gillon}
  et~al.}{2012}]{2012A&A...542A...4G}
{Gillon}, M., et~al. 2012, \aap, 542, A4

\bibitem[\protect\citeauthoryear{{Holman} \& {Murray}}{{Holman} \&
  {Murray}}{2005}]{2005Sci...307.1288H}
{Holman}, M.~J.,  \& {Murray}, N.~W. 2005, Science, 307, 1288

\bibitem[\protect\citeauthoryear{{Hoyer}, {Rojo}, \&
  {L{\'o}pez-Morales}}{{Hoyer} et~al.}{2012}]{2012ApJ...748...22H}
{Hoyer}, S., {Rojo}, P.,  \& {L{\'o}pez-Morales}, M. 2012, \apj, 748, 22

\bibitem[\protect\citeauthoryear{{Hrudkov{\'a}} et~al.}{{Hrudkov{\'a}}
  et~al.}{2009}]{2009IAUS..253..446H}
{Hrudkov{\'a}}, M., {Skillen}, I., {Benn}, C., {Pollacco}, D., {Gibson}, N.,
  {Joshi}, Y., {Harmanec}, P.,  \& {Tulloch}, S. 2009, in IAU Symposium, Vol.
  253, IAU Symposium, 446

\bibitem[\protect\citeauthoryear{{Hut}}{{Hut}}{1981}]{1981A&A....99..126H}
{Hut}, P. 1981, \aap, 99, 126

\bibitem[\protect\citeauthoryear{{Ivezi{\'c}} et~al.}{{Ivezi{\'c}}
  et~al.}{2007}]{2007AJ....134..973I}
{Ivezi{\'c}}, {\v Z}., et~al. 2007, \aj, 134, 973

\bibitem[\protect\citeauthoryear{{Jordi}, {Grebel}, \& {Ammon}}{{Jordi}
  et~al.}{2006}]{2006A&A...460..339J}
{Jordi}, K., {Grebel}, E.~K.,  \& {Ammon}, K. 2006, \aap, 460, 339

\bibitem[\protect\citeauthoryear{{Kundurthy} et~al.}{{Kundurthy}
  et~al.}{2011}]{2011ApJ...731..123K}
{Kundurthy}, P., {Agol}, E., {Becker}, A.~C., {Barnes}, R., {Williams}, B.,  \&
  {Mukadam}, A. 2011, \apj, 731, 123

\bibitem[\protect\citeauthoryear{{Kundurthy} et~al.}{{Kundurthy}
  et~al.}{2013a}]{praveen-xo2}
{Kundurthy}, P., {Barnes}, R., {Becker}, A.~C., {Agol}, E., {Williams}, B.,  \&
  {Mukadam}, A. 2013a, \apj, Submitted

\bibitem[\protect\citeauthoryear{{Kundurthy} et~al.}{{Kundurthy}
  et~al.}{2013b}]{praveen-tres3}
{Kundurthy}, P., {Becker}, A.~C., {Agol}, E., {Barnes}, R.,  \& {Williams}, B.
  2013b, \apj, Accepted

\bibitem[\protect\citeauthoryear{{Lendl} et~al.}{{Lendl}
  et~al.}{2012}]{2012arXiv1212.3553L}
{Lendl}, M., {Gillon}, M., {Queloz}, D., {Alonso}, R., {Fumel}, A., {Jehin},
  E.,  \& {Naef}, D. 2012, ArXiv e-prints

\bibitem[\protect\citeauthoryear{{Maxted}, {Koen}, \& {Smalley}}{{Maxted}
  et~al.}{2011}]{2011MNRAS.418.1039M}
{Maxted}, P.~F.~L., {Koen}, C.,  \& {Smalley}, B. 2011, \mnras, 418, 1039

\bibitem[\protect\citeauthoryear{{Mukadam} et~al.}{{Mukadam}
  et~al.}{2011}]{2011PASP..123.1423M}
{Mukadam}, A.~S., {Owen}, R., {Mannery}, E., {MacDonald}, N., {Williams}, B.,
  {Stauffer}, F.,  \& {Miller}, C. 2011, \pasp, 123, 1423

\bibitem[\protect\citeauthoryear{{Rodgers} et~al.}{{Rodgers}
  et~al.}{2006}]{2006AJ....132..989R}
{Rodgers}, C.~T., {Canterna}, R., {Smith}, J.~A., {Pierce}, M.~J.,  \&
  {Tucker}, D.~L. 2006, \aj, 132, 989

\bibitem[\protect\citeauthoryear{{Sada} et~al.}{{Sada}
  et~al.}{2012}]{2012PASP..124..212S}
{Sada}, P.~V., et~al. 2012, \pasp, 124, 212

\bibitem[\protect\citeauthoryear{{Southworth} et~al.}{{Southworth}
  et~al.}{2009}]{2009MNRAS.396.1023S}
{Southworth}, J., et~al. 2009, \mnras, 396, 1023

\bibitem[\protect\citeauthoryear{{Southworth} et~al.}{{Southworth}
  et~al.}{2010}]{2010MNRAS.408.1680S}
{Southworth}, J., et~al. 2010, \mnras, 408, 1680

\bibitem[\protect\citeauthoryear{{Steffen} et~al.}{{Steffen}
  et~al.}{2012a}]{2012MNRAS.421.2342S}
{Steffen}, J.~H., et~al. 2012a, \mnras, 421, 2342

\bibitem[\protect\citeauthoryear{{Steffen} et~al.}{{Steffen}
  et~al.}{2012b}]{2012PNAS..109.7982S}
{Steffen}, J.~H., et~al. 2012b, Proceedings of the National Academy of Science,
  109, 7982

\bibitem[\protect\citeauthoryear{{Stubbs} et~al.}{{Stubbs}
  et~al.}{2007}]{2007PASP..119.1163S}
{Stubbs}, C.~W., et~al. 2007, \pasp, 119, 1163

\bibitem[\protect\citeauthoryear{{Tegmark} et~al.}{{Tegmark}
  et~al.}{2004}]{2004PhRvD..69j3501T}
{Tegmark}, M., et~al. 2004, \prd, 69, 103501

\bibitem[\protect\citeauthoryear{{Tingley} et~al.}{{Tingley}
  et~al.}{2011}]{2011A&A...536L...9T}
{Tingley}, B., et~al. 2011, \aap, 536, L9

\bibitem[\protect\citeauthoryear{{Tregloan-Reed} \&
  {Southworth}}{{Tregloan-Reed} \& {Southworth}}{2012}]{2012arXiv1212.0686T}
{Tregloan-Reed}, J.,  \& {Southworth}, J. 2012, ArXiv e-prints

\bibitem[\protect\citeauthoryear{{Triaud} et~al.}{{Triaud}
  et~al.}{2010}]{2010A&A...524A..25T}
{Triaud}, A.~H.~M.~J., et~al. 2010, \aap, 524, A25

\bibitem[\protect\citeauthoryear{{Winn} et~al.}{{Winn}
  et~al.}{2009}]{2009AJ....137.3826W}
{Winn}, J.~N., {Holman}, M.~J., {Carter}, J.~A., {Torres}, G., {Osip}, D.~J.,
  \& {Beatty}, T. 2009, \aj, 137, 3826

\bibitem[\protect\citeauthoryear{{Young} et~al.}{{Young}
  et~al.}{1991}]{1991PASP..103..221Y}
{Young}, A.~T., et~al. 1991, \pasp, 103, 221

\bibitem[\protect\citeauthoryear{{Zechmeister} \& {K{\"u}rster}}{{Zechmeister}
  \& {K{\"u}rster}}{2009}]{2009A&A...496..577Z}
{Zechmeister}, M.,  \& {K{\"u}rster}, M. 2009, \aap, 496, 577

\end{thebibliography}

\end{document}